\documentclass[conference]{IEEEtran}
\IEEEoverridecommandlockouts
\usepackage{cite}
\usepackage{amsmath,amssymb,amsfonts}
\usepackage{graphicx}
\usepackage{textcomp}

\usepackage{algorithm}
\usepackage{algpseudocode}

\usepackage{graphicx,xfp}
\usepackage{multicol}
\usepackage{multirow}
\usepackage{color}
\usepackage{xcolor}
\usepackage{booktabs}
\usepackage{graphicx}
\usepackage{nicefrac}
\usepackage{amsmath}

\usepackage{array}
\newcolumntype{P}[1]{>{\centering\arraybackslash}p{#1}}

\usepackage{graphicx}
\usepackage{fancyhdr}
\usepackage{amssymb}
\usepackage{url}
\usepackage{color}
\usepackage{tikz}
\usetikzlibrary{positioning}
\usepackage{pgfplots}
\usepackage{marvosym}

\usepackage{float}
\usepackage{todonotes}
\usepackage{comment}
\usepackage{subcaption}

\usepackage{hyperref}

\usepackage{tikz}
\usepackage{tikzscale}
\usepackage{xspace}
\usepackage{pgfplots}
\usepgfplotslibrary{groupplots}
\pgfplotsset{compat=1.16}
\usepackage{xparse}
\NewDocumentCommand{\Log}{o}{%
  \IfNoValueTF{#1}{}{{}^{#1}\!}\log}%

\usetikzlibrary{external}
\newtheorem{remark}{Remark}%

\DeclareMathOperator{\arccosh}{arcCosh}

\usepackage{xcolor}
\def\BibTeX{{\rm B\kern-.05em{\sc i\kern-.025em b}\kern-.08em
    T\kern-.1667em\lower.7ex\hbox{E}\kern-.125emX}} 

\begin{document}

\title{PCD2Vec: A \textbf{P}oisson \textbf{C}orrection
  \textbf{D}istance Based Approach for Viral Host Classification}


\author{\IEEEauthorblockN{Sarwan Ali$^*$, Taslim Murad$^*$, and Murray Patterson}
\IEEEauthorblockA{\textit{Department of Computer Science, Georgia State University},
Atlanta, USA \\
\{sali85, tmurad2\}@student.gsu.edu, mpatterson30@gsu.edu \\
$^*$ Equal Contribution
}
}
\maketitle


\begin{abstract}
  Coronaviruses are membrane-enveloped, non-segmented positive-strand
  RNA viruses belonging to the Coronaviridae family.  They are
  primarily divided into two subfamilies, Letovirinae and
  Coronavirinae, with the majority of these viruses belonging to
  the latter subfamily.  Various animal species, mainly mammalian and
  avian, are severely infected by various coronaviruses, causing
  serious concerns like the recent pandemic (COVID-19) --- one example
  of the impact of these viruses on human health as well as the global
  economy.  Therefore, building a deeper understanding of these
  viruses is essential to devise prevention and mitigation mechanisms.
  Coronaviruses have an invariant genome organization of
  $\approx$30KB, divided into regions that code for non-structural
  and structural proteins.  Among these, an essential structural
  region is the spike region and its resulting protein which is
  responsible for attaching the virus to the host cell membrane.
  Therefore, the usage of only the spike protein, instead of the full
  genome, provides most of the essential information for performing
  analyses such as host classification.
  In this paper, we propose a novel method for predicting the host
  specificity of coronaviruses by analyzing spike protein sequences
  from different viral subgenera and species.  Our method involves
  using the Poisson correction distance to generate a distance matrix,
  followed by using a radial basis function (RBF) kernel and kernel
  principal component analysis (PCA) to generate a low-dimensional
  embedding.  Finally, we apply classification algorithms to the
  low-dimensional embedding to generate the resulting predictions of
  the host specificity of coronaviruses.  We provide theoretical
  proofs for the non-negativity, symmetry, and triangle inequality
  properties of the Poisson correction distance metric, which are
  important properties in a machine-learning setting.  By encoding the
  spike protein structure and sequences using this comprehensive
  approach, we aim to uncover hidden patterns in the biological
  sequences to make accurate predictions about host specificity.
  Finally, our classification results illustrate that our method can
  achieve higher predictive accuracy and improve performance over
  existing baselines.
\end{abstract}

\begin{IEEEkeywords}
  Host Classification, Spike Sequence, Coronaviruses, Sequence
  Analysis, Classification
\end{IEEEkeywords}

\section{Introduction}

Coronaviruses are well-known for causing pandemics throughout
history. For instance, in 2003, a SARS-CoV (severe acute respiratory
syndrome coronavirus) pandemic originated in China, probably from
bats, spread worldwide, and caused 8,000 cases and 800
deaths~\cite{CDC_2003}. Then, in 2012, a MERS-CoV (Middle East
respiratory syndrome coronavirus) was reported in Saudi Arabia, spread
from camels to humans, causing 2,500 cases and 800
deaths~\cite{WHO_2012_MERS-CoV}. Finally, in 2019, a SARS-CoV-2
originated in China and causing the COVID-19 pandemic, with
102,171,644 cases and 1,103,615 deaths reported so far only in the
United States by CDC (Centers for Disease Control and Prevention).

\begin{figure}[h!]
  \centering \includegraphics[scale = 0.23] {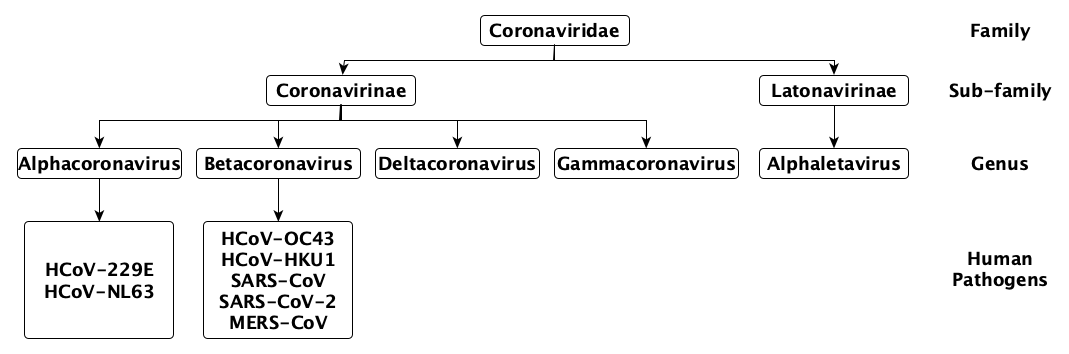}
  \caption{Taxonomic classification scheme for different
    Coronaviridae.}
  \label{fig_covid_family_tree}
\end{figure}

Moreover, coronaviruses have recently affected many living organisms,
including humans, animals, and birds. Coronaviruses are positive-sense
enveloped RNA viruses consisting of two major subfamilies, namely
Coronavirinae and Letovirinae, comprising five genera, namely
Alphacoronavirus, Betacoronavirus, Gammacoronavirus, Deltacoronavirus,
and Alphaletovirus. We can also divide certain major lineages of
coronaviruses, such as SARS-CoV-2 --- the virus responsible for the
COVID-19 pandemic --- into different variant categories such as Alpha,
Beta, Delta, Omicron, etc.  The taxonomy of the major groups
within the Coronaviridae family is given in
Figure~\ref{fig_covid_family_tree}

\begin{figure}[h!]
  \centering \includegraphics[scale = 0.3] {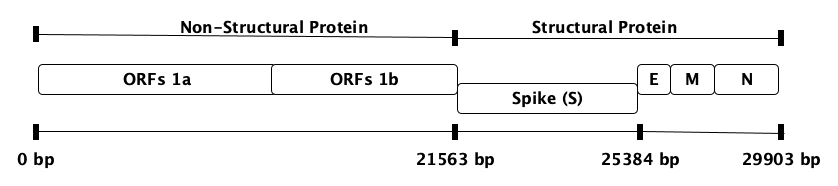}
  \caption{The SARS-CoV-2 genome consists of a structural (S, E, M, N)
    and a non-structural (ORF1ab) part, forming a length of almost
    $30K$ bp. Among them, the spike (S) region is used to attach the
    virus to the host cell membrane, and it also contains the major
    mutations related to viral transmissibility. This S region is used
    to do further host classification.}
  \label{fig_spike_seq}
\end{figure}

Here, our main research question is \textit{``based on a coronavirus
  sequence, can we train a model to perform efficient host
  classification?''}. Since we have dozens of hosts in our data, it
is a multi-class classification problem. As the full genome sequence
(of $\approx$30KB) of the virus is long, it is more efficient to
extract the most useful part of this sequence (the region having the
largest effect on the host classification) to do further analysis, and
for this purpose, we only use the spike region of the virus, or more
accurately, the protein (amino acid sequence) that it produces, as it
is most important for host specificity~\cite{kuzmin2020machine}.  This
is because virus-cell interaction is done through binding the spike
protein to the ACE2 cell receptor, a key step prior to virus-cell
membrane fusion and viral entry into the host cell. As a result, a
disproportionate number of mutations (with respect to the length of
the genome) occur in the spike region.  Therefore, we extract only the
spike protein sequences from the raw data and used them to perform
host classification. A detailed structure of the SARS-CoV-2 genome can be
seen in Figure~\ref{fig_spike_seq}.

Various methods to predict the host specificity of coronaviruses have
been applied so far. Some of them use complete viral genomes while
others prefer individual genes to do the analysis. For instance,
~\cite{kuzmin2020machine} reported the use of four different
machine-learning methods to analyze spike protein sequences from
several coronaviruses. The study's major conclusion was that the host
specificity of coronaviruses could be predicted with accuracy using
only the spike protein sequences. 

In this work, we formulate the method of host classification of
coronaviruses by using the concept of Poisson Correction Distance
(PCD). PCD is a measure of the difference in amino acid composition
between two protein sequences and is specifically designed for the
analysis of biological sequences. The theoretical basis for this
distance measure is the Poisson distribution, which models the number
of events occurring in a fixed interval of time. The PCD formula uses
the observed and expected frequencies of each amino acid in two
sequences and the Poisson distribution to calculate the distance
between the sequences.  This distance is a good measure because it
takes into account both the observed and expected frequencies of each
amino acid in the sequences, and it also considers the variability in
the frequencies of the amino acids by using the Poisson
distribution. Therefore, given the spike protein sequences of
SARS-CoV-2 virus, we follow PCD to classify the sequences into their
respective hosts using machine learning (ML) models. However, before
the classification step, we reduce the distance matrix dimensionality
by employing RBF kernel and kernel PCA. As RBF kernel is known to
perform well in non-linear classification tasks and is often used in
the analysis of biological sequences. Similarly, kernel PCA allows us
to perform dimensionality reduction while preserving the non-linear
structure of the data, which can lead to better separation between
different classes and improve classification
performance. Additionally, our method retains the maximum information
for classification because it operates on raw spike sequences, instead
of any prior embeddings generated out of the sequences.

Our contributions to this paper are as follows:
\begin{enumerate}
\item \textbf{Efficient Prediction:} We show that coronavirus hosts
  can be efficiently predicted using spike sequences only.
\item \textbf{Incorporation of biological knowledge:} Our method to
  generate a low-dimensional embedding, based on the Poisson
  correction distance (PCD), better captures the biological
  relationships between the spike protein sequences in the
  classification task, which general distance measures /
  representation learning methods may not consider.
\item \textbf{Use of RBF kernel:} We used the RBF kernel to project
  the data into high dimensional space, which has been proven to
  perform well in non-linear classification tasks and is often used in
  the analysis of biological sequences.
\item \textbf{Use of kernel PCA:} We used Kernel PCA, which allows us
  to perform dimensionality reduction while preserving the non-linear
  structure of the data. This can lead to better separation between
  the different classes and improved classification performance.
\item \textbf{Theoretical proofs for three properties:} We provide
  theoretical proofs for the triangle inequality, symmetry, and
  non-negativity properties to ensure the validity of the distance
  metric used in our method, which can add confidence to the results
  obtained from our method.
\end{enumerate}

The rest of the paper is organized as follows:
Section~\ref{sec_related_work} contains the detailed literature review of the
paper. The proposed approach is described in
Section~\ref{sec_proposed_approach}. Section~\ref{sec_exp} discusses the data source, evaluation metrics, and baseline methods used for performing the experiments.
The results are reported in
Section~\ref{sec_results}. Finally, we conclude the paper in
Section~\ref{sec_conclusion}.

\section{Literature Review} \label{sec_related_work}
The availability of a huge amount of COVID-19 data has enabled researchers to put more effort into analyzing the behavior of the Coronavirus. The motivation behind this kind of study is to build prevention mechanisms, like vaccines, etc, and control the spread of the virus. Sequence analysis~\cite{krishnan2021predicting,heinzinger2019modeling} is a popular approach used to conduct such studies, especially in the bioinformatics domain. For instance, the authors in ~\cite{kuzmin2020machine} used different ML algorithms to study the spike sequences by converting the sequences to one-hot encoded vectors for the purpose of predicting the host specificity of Coronavirus. However, this technique undergoes sparsity and a curse of dimensionality challenges. Likewise,~\cite{tuli2020predicting} used Generalized Inverse Wei-bull distribution to predict the growth and trend of the COVID-19 pandemic. In~\cite{ahmad2020number} they have compared various research papers that are using machine learning-based approaches for predicting confirmed cases of COVID-19 worldwide. 

Furthermore, more traditional methods for analyzing sequencing data typically
employ a phylogenetic approach~\cite{doolittle1999phylogenetic,stevens2003phylogenetics,minh_2020_iqtree2}.
However, the number of sequences currently available for viruses, such
as SARS-CoV-2, is several orders of magnitude which is beyond the
capabilities of the traditional methods, making ML approaches a more attractive alternative.

Another category of sequence investigation strategies is based on $k$-mers~\cite{ali2021k,tayebi2021robust} (sub-strings of length $k$). The use of $k$-mer frequencies for phylogenetic applications was first explored in~\cite{Blaisdell1986AMeasureOfSimilarity}, which proposed constructing accurate phylogenetic trees from several coding and non-coding nDNA sequences. Similarly, authors in~\cite{ali2021spike2vec} provided an alignment-free $k$-mers-based approach for classifying SARS-CoV-2 sequences. PWKmer~\cite{ma2020phylogenetic} technique performs phylogenetic analysis of HIV-1 viruses by utilizing $k$-mers counts and position distribution information. Although $k$-mers-based approaches achieve reasonable performance, their applications are inherently limited. As $k$ increases and quickly approaches a sparse matrix of (binary) counts, the likelihood of seeing any particular $k$-mer decreases~\cite{singh2017gakco}. Moreover, any statistical method that uses $k$-mers is vulnerable to inaccurate $k$-mer frequency estimates. 
Another domain of research involves using kernel methods for sequence classification~\cite{farhan2017efficient,ali2022efficient}. However, most of these methods only work with traditional $k$-mers spectrum, which may not capture all (hidden) biological patterns from the sequences, and hence could under-perform.

\section{Proposed Approach} \label{sec_proposed_approach}
This section discusses our proposed approach, Poisson Correction Distance Based Embedding (PCD2Vec), to perform host classification of Coronavirus based on the virus's spike sequences. It also highlights the theoretical proofs of the triangle inequality, symmetry, and non-negativity properties of the distance matrix used by our method.

The pseudo-code for PCD2Vec is illustrated in Algorithm~\ref{algo_pcd}. We can observe that given the set of spike protein sequences, it returns the embeddings, which are further utilized for doing the classification. The algorithm starts with initializing a distance matrix by zeros, as shown in line 1. Then we are computing only the upper triangular values of this matrix (from line $2$-$23$), as it is a symmetric matrix with diagonal elements being equal to zero. Each element in the matrix corresponds to a distance value between the pair of sequences. This distance value is calculated first by counting the occurrences of each amino acid for both sequences respectively to get observed frequency vectors ($obs\_freq1$ and $obs\_freq1$), as shown in lines $5$-$6$. Then an expected frequency ($exp\_freq$) vector is obtained by averaging the observed frequencies in line $8$. To eliminate any zero value in the observed frequencies, $0.0001$ is added to them. After that, the distance $d$ between the sequence pair is calculated using the PCD formula (as illustrated in lines $10$-$19$), and the formula is as follows,
\begin{equation}
    \begin{aligned}
        d = 2 \times exp\_freq \times (\ln(\frac{obs\_freq1}{exp\_freq})
            + \ln(\frac{obs\_freq2}{exp\_freq}) )
    \end{aligned}
\end{equation}
Once the distance values are computed, the distance matrix is updated with these values at their respective positions, as given in lines $20$-$21$. For $n$ number of sequences, we get an $n$x$n$ symmetric distance matrix. We then pass this distance matrix to RBF kernel having $\sigma = 0.5$ to get the kernel matrix (line $24$). This kernel matrix is further given to kernel PCA to reduce its dimensions to $500$ components (line $25$) and it outputs the final embeddings of the given sequences.

\begin{algorithm}[h!]
\caption{The algorithm for PCD-based embedding generation for spike sequences.}
\label{algo_pcd}
\begin{algorithmic}[1]
\Statex \textbf{Input}: Set of Spike Sequences ($seqs$)
\Statex \textbf{Output}: Embeddings
\State $distances$ $\leftarrow$ $zeros(len(seqs), len(seqs))$
\For{$i$ in $\vert seqs \vert -1$}
    \For{$j$ in $(i+1,\vert seqs \vert)$} \Comment{Upper Triangle Only}
        \State /* Compute the observed frequencies of each amino acid  */
        \State $obs\_freq1$ $\leftarrow$ \Call{AminoAcidFreq}{$seqs[i]$}
        \State $obs\_freq2$ $\leftarrow$ \Call{AminoAcidFreq}{$seqs[j]$}
        \State /* Compute the expected frequencies  */
        \State $exp\_freq$ $\leftarrow$ $ 0.5 \times (obs\_freq1 + obs\_freq2) $
        \State $d$ $\leftarrow$ 0
        \For{$k$ in $\vert 20 \vert$} \Comment{$20$ Amino Acids}
            \If{$exp\_freq[k]$ $>$ 0}
                \State $\epsilon \gets  0.0001$ \Comment{to avoid divided by 0 error}
                \State $obs\_freq1[k]$ $\leftarrow$ $obs\_freq1[k] + \epsilon $
                \State $obs\_freq2[k]$ $\leftarrow$ $obs\_freq2[k] + \epsilon $
                \State Freq\_1 $\leftarrow$ $ln(\frac{obs\_freq1[k]}{exp\_freq[k]})$
                \State Freq\_2 $\leftarrow$ $ln(\frac{obs\_freq2[k]}{exp\_freq[k]})$
                \State Freq $\leftarrow$ Freq\_1 + Freq\_2
                \State $d$ $\leftarrow$  d + 2 $\times$ exp\_freq[k] $\times$ Freq
            \EndIf
        \EndFor
        \State $distances[i, j]$ $\leftarrow$ $d$
        \State $distances[j, i]$ $\leftarrow$ $d$ 
    \EndFor
\EndFor
\State $kernelMatrix$ $\leftarrow$ \Call{RbfKernel}{distances}
\State $Embedding$ $\leftarrow$ \Call{KernelPCA}{$kernelMatrix$}
\end{algorithmic}
\end{algorithm}

\begin{figure}[h!]
  \centering \includegraphics[scale = 0.3] {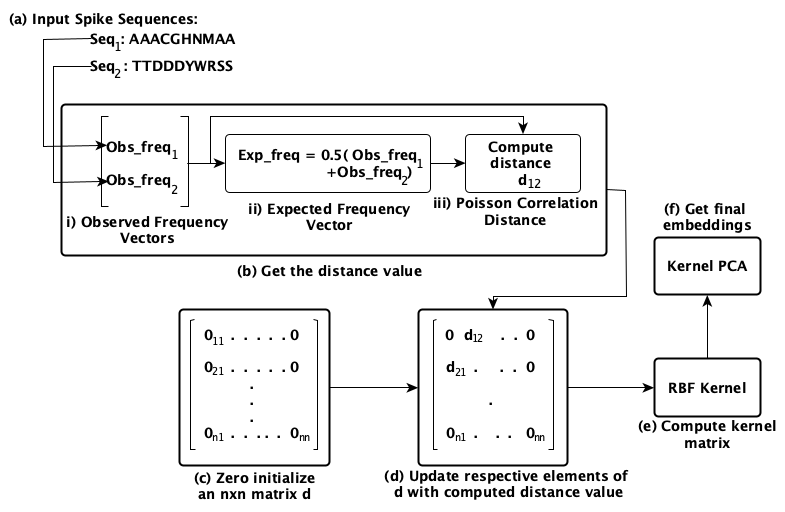}
  \caption{The workflow of PCD2Vec. Given a set of spike sequences, it generates the corresponding numerical embeddings using the Poisson correlation distance, RBF kernel and kernel PCA (steps $a$-$f$). 
  }
  \label{fig_pcd2vec_workflow}
\end{figure}

Moreover, the workflow of PCD2Vec is illustrated in Figure~\ref{fig_pcd2vec_workflow}. Given a set of spike sequences, it iteratively follows the step ($a$-$b$) to get the respective distance values for each pair of sequences. These values are used to update the distance matrix in ($d$). This updated distance matrix is given to RBF kernel to get a kernel matrix in ($e$), and then the final embeddings are generated by passing the kernel matrix to kernel PCA in ($f$).

Furthermore, the details of theoretical proofs of certain properties of the used distance metric are given as follows:

\subsection{Theoretical Properties}
The distance matrix generated using Poison Correction Distance should hold three properties namely ``triangle inequality," ``symmetry," and ``non-negativity". These properties are important for a distance matrix in a machine-learning setting for several reasons:

\begin{enumerate}
    \item Triangle inequality: The triangle inequality property ensures that the distance between two points via a third point is always equal to or greater than the direct distance between the two points. This property ensures that the distance metric is consistent and well-defined.
    \item Symmetry: The symmetry property ensures that the distance between two points is the same regardless of the order in which the points are considered. This property ensures that the distance metric is consistent and unbiased.
    \item Non-negativity: The non-negativity property ensures that the distance between any two points is always non-negative, i.e., it is either zero or a positive number. This property ensures that the distance metric is well-defined and that it has a clear meaning.
\end{enumerate}

These properties ensure that the distance matrix generated is a valid distance metric that can be used for various machine-learning tasks, such as clustering, classification, and dimensionality reduction. The proofs for these properties are given below.

\subsubsection{Triangle inequality}
Let $d(i,j)$ be the distance between two protein sequences $i$ and $j$, and let $d(i,k)$ be the distance between sequences $i$ and $k$. The triangle inequality states that $d(i,j) + d(j,k) \geq d(i,k)$ for any three sequences $i$, $j$, and $k$.

\textbf{Proof:}
The triangle inequality for the Poisson correction distance matrix can be shown as follows:

\begin{equation}
    \begin{aligned}
    d(i,j) + d(j,k) = 2 \times \Sigma(
    \text{exp\_freq}[m] \times \\ (
    \ln (\frac{ \text{obs\_freq1}[m]}{\text{exp\_freq}[m]}) + 
    \ln (\frac{\text{obs\_freq2}[m]}{\text{exp\_freq}[m]})
    )
)
    \end{aligned}
\end{equation}
where $\text{obs\_freq1}$ and $ \text{obs\_freq2}$ are the observed frequencies of amino acids in sequences $i$ and $j$, and $\text{exp\_freq}$ is the average of $\text{obs\_freq1}$ and $\text{obs\_freq2}$.

\begin{equation}
    \begin{aligned}
    d(i,k) = 2 \times \Sigma(\text{exp\_freq}[m] \times \\ (
    \ln (\frac{\text{obs\_freq1}[m]}{\text{exp\_freq}[m]}) + \ln( \frac{\text{obs\_freq3}[m]}{\text{exp\_freq}[m])})))
    \end{aligned}
\end{equation}

where $\text{observed\_freq1}$ and $\text{observed\_freq3}$ are the observed frequencies of amino acids in sequences $i$ and $k$, and $\text{expected\_freq}$ is the average of $\text{observed\_freq1}$ and $\text{observed\_freq3}$.

Since $\ln(x)$ is a strictly increasing function, it follows that $d(i,j) + d(j,k) \geq d(i,k)$ for any three sequences $i$, $j$, and $k$.

\subsubsection{Symmetry}
Let $d(i,j)$ be the distance between two protein sequences $i$ and $j$, then $d(i,j) = d(j,i)$ for any two sequences $i$ and $j$.

\textbf{Proof:}
Since the Poisson correction distance formula is symmetric with respect to the observed frequencies of the two sequences, it follows that $d(i,j) = d(j,i)$ for any two sequences $i$ and $j$.

\subsubsection{Non-negativity}
Let $d(i,j)$ be the distance between two protein sequences $i$ and $j$, then $d(i,j) \geq 0$ for any two sequences $i$ and $j$.

\textbf{Proof:}
The Poisson correction distance formula is based on the logarithm of the ratio of observed and expected frequencies. Since the logarithm of a positive number is always positive, it follows that the Poisson correction distance is always non-negative.

In conclusion, the Poisson correction distance matrix generated in Algorithm~\ref{algo_pcd} satisfies the triangle inequality, symmetry, and non-negativity properties, which are important properties for a distance matrix in a machine learning setting.

\section{Experimental Evaluation} \label{sec_exp}
In this section, we present the experimental evaluation of our proposed method to predict the host specificity of coronaviruses using spike protein sequences. We begin by describing the datasets used in our experiments, including the statistics of the datasets. Then, we explain the feature selection methods used to extract relevant information from the spike protein sequences. Finally, we outline the machine learning classification algorithms used to predict the host specificity of coronaviruses and compare their performance in terms of average accuracy, precision, recall, weighted F1-score, macro F1 score, and ROC-AUC. Since our dataset is imbalanced, so reporting performance using various metrics will provide more meaningful insights to it therefore we are employing 5 different evaluation metrics.
Experiments are performed on a Core i5 system with $32$ GB memory and Windows 10 OS with a 2.4 GHz processor. Our code is implemented in Python and is available online~\footnote{\url{https://github.com/sarwanpasha/PCD2Vec}}.

\subsection{Dataset Analysis}
The protein sequences ($2576$ sequences in total) from different species of mammalian and avian coronaviruses were obtained from the Virus Pathogen Database~\footnote{\url{https://www.viprbrc.org/}}. Amino acid sequences from each host were stored in a different fasta file (one fasta file per host).
Multiple sequence alignments were performed using Clustal Omega~\footnote{\url{https://www.ebi.ac.uk/Tools/msa/clustalo/}}. The distribution of different hosts is given in Table~\ref{tab_data_Stats}.

We can observe that $6$ hosts are dominating while the presence of other hosts is very low. Initially, there was also a problem with some missing information in the sequences. Those sequences were discarded. We selected the sequences having one of the first $6$ hosts (humans, swine, chicken, camel, bat, cat) for the classification and analysis part. The rest of the $7$ hosts were assigned a new class ``others," because their presence is very low in the data (i.e., $\leq 10$ sequences for each of these $7$ hosts). 

\begin{table}[h!]
    \centering
    \begin{tabular}{lc|lc}
    \toprule
    Host & Count & Host & Count \\
    \midrule \midrule
        human & 957  & pangolin & 5 \\
        swine & 785  & duck & 3 \\
        chicken & 309 & chimpanzee & 3 \\
        camel & 265 & goose & 2 \\
        bat & 181 & beluga Whale & 2 \\
        cat & 57 & falcon & 1 \\
        civet & 5 & - & - \\
        \midrule \midrule
        Total  & 2575 \\
        \bottomrule
    \end{tabular}
    \caption{Host (class label) distribution in data.}
    \label{tab_data_Stats}
\end{table}

To use the data for classification, we have to convert the sequences into feature vectors (in euclidean space). For this purpose, we used one hot encoding approach. The overall size of the one-hot encoding vector for each sequence is $24700$. Since this one-hot encoding vector is very high dimensional, it creates the problem of the ``curse of dimensionality". We used different feature selection methods to deal with this problem, which we will be discussing in the next section.

\subsection{Baseline Methods}
We selected baseline methods from different domains, such as feature engineering (PCA, Lasso Regression, Ridge Regression, OHE), neural network (Auto Encoder), kernel function (string kernel), hyperbolic embedding (Poincaré embedding), and pre-trained language model for proteins (Protein Bert), so that we can test the performance of the proposed PCD2Vec more rigorously.

\subsubsection{One-Hot Encoding (OHE)}
It is an algorithm to generate a fixed-length numerical representation of given sequences~\cite{kuzmin2020machine}. It creates a binary vector for each character of the sequence where this vector is of length equal to total number of possible unique values and the corresponding location of the appeared character is marked $1$ in this binary vector and the rest is $0$. Then all the binary vectors corresponding to every character of the given sequence is concatenated to form the final embedding. 

\subsubsection{One-Hot Encoding + PCA}
This method is proposed in~\cite{kuzmin2020machine} where the authors perform host classification by first designing one-hot encoding-based representation for the protein sequences and then applying Principal Component Analysis (PCA)~\cite{abdi2010principal} to compute low dimensional embedding of the data. 
We selected $40$ principal components out of a total of $24700$ features. The reason for selecting only $40$ components is that most of the information was preserved in these components. For reference, we call this method as PCA.

\subsubsection{Ridge Regression}
It is a famous parameter estimation approach used to address the collinearity problem that arises in multiple linear regression frequently~\cite{mcdonald2009ridge}.
Ridge Regression works by introducing a Bias term. Its goal is to increase the bias to improve the variance (i.e., the model's generalization capability). It changes the slope of the line and tries to make it more horizontal like.
Ridge regression is very useful for feature selection because it gives insights into which independent variables are useless (can reduce the slope close to zero). We can eliminate some of the useless independent variables to reduce the dimensions of the overall dataset. The objective function of ridge regression is the following:
\begin{equation}
    min(\text{Sum of square residuals} + \alpha \times \text{slope}^2)
\end{equation}
where $\alpha \times {slope}^2$ is called penalty terms. After applying ridge regression, we selected $7079$ features out of $24700$.

\subsubsection{Lasso Regression}
Lasso regression works similarly to ridge regression~\cite{ranstam2018lasso}. The only difference is that it can reduce the slope exactly $= 0$ rather than close to zero (which is the case with ridge regression). The objective function of ridge regression is the following
\begin{equation}
    min(\text{Sum of square residuals} + \alpha \times \vert \text{slope}\vert)
\end{equation}
where $\alpha \times \vert {slope}\vert$ is called penalty terms. Note that in the penalty term, the cardinality of the slope is taken rather than the square. This helps to reduce the slope of useless variables exactly equal to zero. After applying lasso regression, we selected $813$ features out of $24700$.

\subsubsection{Autoencoder}
It is an unsupervised artificial neural network-based approach~\cite{xie2016unsupervised} (also called self-supervised approach). 
It performs two operations, namely compression, and reconstruction.
One property of autoencoder is that it ignores the features that are noisy and/or redundant. In this way, only those features are considered  (in low dimensions) that are important. It is also called the self-supervised approach because it uses reconstruction loss to decompress the data. When the loss is minimum, it stops. At that stage, the middle (encoded layer) contains the reduced dimensional information that we can use as input for the classification algorithm. 
The number of dimensions that we selected after applying the autoencoder is $200$.

\subsubsection{Poincaré Embeddings}
The Poincaré embedding~\cite{nickel2017poincare} is a technique for embedding high-dimensional data in hyperbolic space. The goal of embedding is to preserve the geometric structure of the data while reducing its dimensionality. Unlike classical embedding techniques such as PCA, which embeds data in Euclidean space, Poincaré embedding embeds data in a space with non-Euclidean geometry (hyperbolic space), where distances between points are measured using the Poincaré distance.
In our case, the goal of using this method is to map the sequence of amino acids (which are used as input) into a hyperbolic space using the Poincaré distance formula. We start by counting the number of occurrences of each amino acid in the sequence and then normalizing the counts to obtain amino acid frequencies. 
After normalization, We map the normalized frequencies into a hyperbolic space using gradient descent and the Poincaré distance formula. The Poincaré distance formula is as follows:
\begin{equation}
    d(\mathbf{x}, \mathbf{y}) = \arccosh \left( 1 + \frac{2 \left| \mathbf{x} - \mathbf{y} \right|^2}{(1 - \left| \mathbf{x} \right|^2)(1 - \left| \mathbf{y} \right|^2)} \right)
\end{equation}
where $\mathbf{x}$ and $\mathbf{y}$ are points in the hyperbolic space, and $\left| \cdot \right|$ represents the Euclidean norm.
We first initialize the embedding as a random point in the hyperbolic space and sets the learning rate ($0.001$) and the number of iterations ($1000$) for optimization. It then updates the embedding using gradient descent until the desired number of iterations is reached.
The final result (output) of this method is an embedding that represents the input sequence into the  hyperbolic space. We selected $50$ as the length of the final embedding. The hyperparameters such as learning rate, number of iterations, and embedding length are tuned using the standard validation set approach.

\subsubsection{String Kernel}
This method~\cite{ali2022efficient} has proposed a technique to compute similarity between two sequences by using the number of matched and mis-matched $k$-mers. The matches between 2 $k$-mers is determine by having them at a distance $m$. It also claims to provide a computationally efficient closed-form solution for the size of the intersection of d-mismatch neighborhoods. String kernel yields a square kernel matrix of dimensions equal to the length of sequences, and this matrix is given to kernel PCA to reduce its dimensions. We used $k=3$ for our experiments.

\subsubsection{Protein Bert}
Protein Bert~\cite{Brandes2022ProteinBERT} provides a deep language model to analyze protein sequences. This model is pretrained by combining Gene Ontology (GO) annotation prediction task with language modeling. To make the model flexible to deal with long sequences efficiently, new architectural elements are introduced in it. 

\subsection{Classification Task}
For classification, we used the following ML algorithms: Support Vector Machine (SVM), Naive Bayes (NB), Multilayer Perceptron (MLP), k-Nearest Neighbor (k-NN) (where $k = 3$), Random Forest (RF), Logistic Regression (LR). 
For experiments, we used a 10-fold cross-validation approach. We first randomly divide the data into 70-30\% (where 30\% is the held-out set). On the 70\% data, we use 10-fold cross-validation to optimize the hyperparameters. In the end, we validate the results on the 30\% held-out testing set. This 70-30\% split process is repeated 5 times, and we compute averages and variances over these runs and computed results for different performance metrics. The performance metrics include Average accuracy, precision, recall, weighted F1, F1 macro, and ROC AUC macro. One of the reasons for the selection that many performance metrics was to analyze the underlying classifiers in detail. Since we are dealing with a class imbalance problem, reporting only the accuracy does not give us a detailed overview of a classifier's actual performance. 
\begin{remark}
For binary evaluation metrics, we use the one-vs-rest approach.
\end{remark}

\section{Results and Discussion}\label{sec_results}
This section provides insight into the classification performance result obtained by our proposed system and compares it with the other baselines. 

The results in Table~\ref{tbl_pca_accy} show average and variance values for various performance metrics using different baseline methods and our proposed system. For all the results, we can observe that random forest consistently outperforms all other machine learning classifiers. We can also note that the values for variance are very low. 

\subsection{Comparison with feature engineering methods}
Overall, our method is outperforming all the feature engineering-based baselines in terms of every evaluation metric. Like, PCD2Vec achieves $2\%$ average accuracy improvement over PCA, OHE, Ridge Regression and Lasso Regression for RF classifier. As RF model is showcasing the best performance results for all the methods as compared to the other classifiers. 

\subsection{Comparison with Neural network baseline}
We can observe that our method is showing better performance than the NN-based baseline (Autoencoder). For instance, PCD2Vec has $2\%$ accuracy and $1\%$ precision improvement over Autoencoder for RF classifier.

\subsection{Comparison with kernel function based baseline}
The String Kernel technique also exhibits lower performance then PCD2Vec. For example, it has $2\%$ and $1\%$ lower accuracy and precision then PCD2Vec.

\subsection{Comparison with hyperbolic embedding based baseline}
PCD2Vec portrays huge performance improvement over Poincaré Embedding method for all evaluation metrics. Like, it achieves $19\%$ higher accuracy and $23\%$ higher precision then the Poincaré Embedding method for RF classifier.

\subsection{Comparison with pre-trained language model baseline}
Comparing PCD2Vec to a more advance pre-trained language model based-baseline (Protein Bert) illustrates that PCD2Vec is able to maintain higher performance results. For instance, PCD2Vec has $5\%$ accuracy and $4\%$ precision improvement over Protein Bert. A reason for this behavior can be that Protein Bert may not have been trained on a similar dataset as breast cancer therefore it is unable to generalize well on our used breast cancer dataset.

Furthermore, for the majority of the methods and classifiers, the average ROC-AUC values are close to 0.9, indicating that the performance of the models is good in terms of separating the positive and negative classes. The variance of the ROC-AUC values is small, implying that the results are consistent across different runs of the models.

\begin{table*}[h!]
    \centering
    \resizebox{0.99\textwidth}{!}{
    \begin{tabular}{@{\extracolsep{4pt}}p{1.2cm}p{0.8cm}cccccccccccp{0.25cm}}
    \toprule
        \multirow{3}{*}{Method} & \multirow{3}{*}{Classifier} & \multicolumn{2}{c}{Accuracy} & \multicolumn{2}{c}{Precision} & \multicolumn{2}{c}{Recall} & \multicolumn{2}{c}{F1 Weigh.} & \multicolumn{2}{c}{F1 Macro} &  \multicolumn{2}{p{2cm}}{ROC AUC} \\
        \cmidrule{3-4} \cmidrule{5-6} \cmidrule{7-8} \cmidrule{9-10} \cmidrule{11-12} \cmidrule{13-14}
        & & Avg. & Var. & Avg. & Var. & Avg. & Var. & Avg. & Var. & Avg. & Var. & Avg. & Var.\\
        \midrule \midrule
        \multirow{6}{*}{PCA} & SVM & 0.90 & 0.0001 & 0.92 & 0.0002 & 0.90 & 0.0001 & 0.88 & 0.0003 & 0.82 & 0.0002 & 0.90 & 0.0125 \\
        & NB & 0.86 & 0.0002 & 0.93 & 0.0001 & 0.86 & 0.0004 & 0.87 & 0.0002 & 0.77 & 0.0005 &  0.93 & 0.0114 \\
        & MLP & 0.91 & 0.0001 & 0.91 & 0.0001 & 0.91 & 0.0002 & 0.90 & 0.0001 & 0.84 & 0.0004 &  0.90 & 0.0212 \\
        & KNN & 0.94 & 0.0003 & 0.94 & 0.0001 & 0.94 & 0.0003 & 0.93 & 0.0002 & 0.86 & 0.0001 &  0.92 & 0.0152 \\
        & RF & 0.95 & 0.0002 & 0.96 & 0.0001 & 0.95 & 0.0003 & 0.95 & 0.0002 & 0.95 & 0.0002 &  0.97 & 0.0099 \\
        & LR & 0.91 & 0.0003 & 0.91 & 0.0001 & 0.91 & 0.0002 & 0.90 & 0.0001 & 0.84 & 0.0002 &  0.90 & 0.0104 \\
        
        \midrule
        \multirow{6}{*}{AutoEncoder} & SVM & 0.92 & 0.0003 & 0.92 & 0.0003 & 0.92 & 0.0002 & 0.90 & 0.0001 & 0.82 & 0.0002 & 0.89 & 0.0099 \\
        & NB & 0.78 & 0.0001 & 0.87 & 0.0002 & 0.78 & 0.0001 & 0.80 & 0.0001 & 0.69 & 0.0003 &  0.89 & 0.0107 \\
        & MLP & 0.93 & 0.0001 & 0.94 & 0.0003 & 0.93 & 0.0001 & 0.93 & 0.0002 & 0.89 & 0.0002 &  0.94 & 0.0097 \\
        & KNN & 0.93 & 0.0004 & 0.94 & 0.0002 & 0.93 & 0.0004 & 0.93 & 0.0001 & 0.86 & 0.0002 &  0.93 & 0.0092 \\
        & RF & 0.95 & 0.0002 & 0.96 & 0.0001 & 0.95 & 0.0002 & 0.95 & 0.0003 & 0.94 & 0.0001 &  0.97 & 0.0114 \\
        & LR & 0.91 & 0.0002 & 0.91 & 0.0004 & 0.91 & 0.0002 & 0.90 & 0.0002 & 0.80 & 0.0004 &  0.88 & 0.0012 \\
        \midrule
        \multirow{6}{0.9cm}{Lasso Regression} & SVM & 0.95 & 0.0003 & 0.96 & 0.0004 & 0.95 & 0.0003 & 0.95 & 0.0001 & 0.94 & 0.0002 & 0.97 & 0.0094 \\
        & NB & 0.92 & 0.0001 & 0.95 & 0.0005 & 0.92 & 0.0003 & 0.92 & 0.0004 & 0.91 & 0.0005 &  0.96 & 0.0059 \\
        & MLP & 0.94 & 0.0003 & 0.95 & 0.0005 & 0.94 & 0.0002 & 0.94 & 0.0004 & 0.90 & 0.0005 &  0.94 & 0.0098 \\
        & KNN & 0.93 & 0.0001 & 0.92 & 0.0003 & 0.93 & 0.0002 & 0.92 & 0.0003 & 0.88 & 0.0002 &  0.92 & 0.0047 \\
        & RF & 0.95 & 0.0001 & 0.96 & 0.0001 & 0.95 & 0.0002 & 0.95 & 0.0001 & 0.95 & 0.0003 & 0.97 & 0.0098 \\
        & LR & 0.94 & 0.0003 & 0.94 & 0.0004 & 0.94 & 0.0003 & 0.93 & 0.0001 & 0.91 & 0.0002 &  0.94 & 0.0025 \\
        \midrule
        \multirow{6}{0.9cm}{Ridge Regression} & SVM & 0.95 & 0.0001 & 0.96 & 0.0002 & 0.95 & 0.0001 & 0.95 & 0.0005 & 0.94 & 0.0004 & 0.97 & 0.0075 \\
        & NB & 0.94 & 0.0004 & 0.96 & 0.0005 & 0.94 & 0.0001 & 0.94 & 0.0002 & 0.92 & 0.0005 &  0.96 & 0.0071 \\
        & MLP & 0.93 & 0.0002 & 0.94 & 0.0003 & 0.93 & 0.0002 & 0.93 & 0.0005 & 0.88 & 0.0006 &  0.93 & 0.0064 \\
        & KNN & 0.92 & 0.0001 & 0.92 & 0.0002 & 0.92 & 0.0001 & 0.92 & 0.0005 & 0.86 & 0.0001 &  0.91 & 0.0027 \\
        & RF & 0.95 & 0.0002 & 0.96 & 0.0004 & 0.95 & 0.0002 & 0.95 & 0.0003 & 0.94 & 0.0005 &  0.97 & 0.0028 \\
        & LR & 0.94 & 0.0003 & 0.94 & 0.0002 & 0.94 & 0.0003 & 0.94 & 0.0001 & 0.91 & 0.0002 &  0.95 & 0.0046 \\
        \midrule
        \multirow{6}{1.5cm}{OHE} & SVM & 0.95 & 0.0001 & 0.96 & 0.0003 & 0.95 & 0.0002 & 0.95 & 0.0005 & 0.94 & 0.0004 & 0.97 & 0.0078 \\
        & NB & 0.94 & 0.0002 & 0.96 & 0.0001 & 0.94 & 0.0002 & 0.94 & 0.0005 & 0.93 &  0.0001 &  0.97 &  0.0051 \\
        & MLP & 0.94 & 0.0003 & 0.94 & 0.0002 & 0.94 & 0.0003 & 0.93 & 0.0004 & 0.89 & 0.0001 &  0.94 & 0.0069 \\
        & KNN & 0.93 & 0.0001 & 0.95 & 0.0003  & 0.93 & 0.0002 & 0.93 & 0.0001  & 0.90 & 0.0004 &  0.95 & 0.0074 \\
        & RF & 0.95 & 0.0004 & 0.96 & 0.0002 & 0.95 & 0.0004 & 0.95 & 0.0003 & 0.94 & 0.0001 & 0.97 & 0.0028\\
        & LR & 0.94 & 0.0002 & 0.95 & 0.0004 & 0.94 & 0.0002 & 0.94 & 0.0001 & 0.93 & 0.0003 & 0.96 & 0.0088 \\
        \midrule
        \multirow{6}{1.5cm}{String Kernel} 
        & SVM & 0.94 & 0.0007 & 0.95 & 0.0002 & 0.94 & 0.0007 & 0.94 & 0.0014 & 0.90 & 0.0006 & 0.95 & 0.0019 \\
        & NB & 0.69 & 0.0019 & 0.86 & 0.0017 & 0.69 & 0.0019 & 0.72 & 0.0011 & 0.70 & 0.0019 & 0.86 & 0.0004 \\
        & MLP & 0.82 & 0.0011 & 0.81 & 0.0030 & 0.82 & 0.0031 & 0.81 & 0.0025 & 0.44 & 0.0040 & 0.71 & 0.0023 \\
        & KNN & 0.93 & 0.0007 & 0.93 & 0.0055 & 0.93 & 0.0097 & 0.92 & 0.0092 & 0.61 & 0.0023 & 0.82 & 0.0030 \\
        & RF & 0.95 & 0.0010 & 0.96 & 0.0025 & 0.95 & 0.0010 & 0.95 & 0.0063 & 0.91 & 0.0059 & 0.95 & 0.0083 \\
        & LR & 0.94 & 0.0008 & 0.95 & 0.0017 & 0.94 & 0.0071 & 0.94 & 0.0018 & 0.90 & 0.0067 & 0.95 & 0.0015 \\
        \midrule
        \multirow{6}{1.5cm}{Poincaré Embedding} 
         & SVM & 0.39 & 0.0001 & 0.34 & 0.0001 & 0.39 & 0.0001 & 0.33 & 0.0001 & 0.10 & 0.0008 & 0.52 & 0.0004 \\
 & NB & 0.74 & 0.0001 & 0.71 & 0.0008 & 0.74 & 0.0001 & 0.72 & 0.0004 & 0.30 & 0.0022 & 0.64 & 0.0005 \\
 & MLP & 0.64 & 0.0001 & 0.56 & 0.0002 & 0.64 & 0.0001 & 0.59 & 0.0002 & 0.21 & 0.0008 & 0.58 & 0.0001 \\
 & KNN & 0.60 & 0.0002 & 0.57 & 0.0002 & 0.60 & 0.0002 & 0.57 & 0.0003 & 0.21 & 0.0005 & 0.58 & 0.0001 \\
 & RF & 0.78 & 0.0005 & 0.74 & 0.0009 & 0.78 & 0.0005 & 0.73 & 0.0008 & 0.31 & 0.0019 & 0.64 & 0.0004 \\
 & LR & 0.34 & 0.0003 & 0.30 & 0.0013 & 0.34 & 0.0003 & 0.27 & 0.0003 & 0.08 & 0.0001 & 0.50 & 0.0000 \\
        \midrule
        \multirow{1}{1.5cm}{Protein Bert} & 
        - & 0.92 & 0.0004 & 0.93 & 0.0002 & 0.92 & 0.0003 & 0.91 & 0.0001 & 0.86 & 0.0002 & 0.92 & 0.0003 \\
        \midrule
        \multirow{6}{1.5cm}{PCD2Vec (Ours)} & 
        SVM & 0.87 & 0.0098 & 0.90 & 0.0508 & 0.87 & 0.0098 & 0.86 & 0.0193 & 0.74 & 0.1030 & 0.87 & 0.0505 \\
        & NB & 0.68 & 0.0470 & 0.87 & 0.0227 & 0.68 & 0.0470 & 0.71 & 0.0450 & 0.75 & 0.0724 & 0.90 & 0.0304 \\
        & MLP & 0.84 & 0.0209 & 0.85 & 0.0219 & 0.84 & 0.0209 & 0.84 & 0.0236 & 0.64 & 0.0839 & 0.79 & 0.0392 \\
        & KNN & 0.93 & 0.0107 & 0.94 & 0.0153 & 0.93 & 0.0107 & 0.93 & 0.0136 & 0.70 & 0.0663 & 0.90 & 0.0430 \\
        & RF & \textbf{0.97} & 0.0085 & \textbf{0.97} & 0.0099 & \textbf{0.96} & 0.0085 & \textbf{0.96} & 0.0090 & \textbf{0.98} & 0.0842 & \textbf{0.99} & 0.0454 \\
        & LR & 0.86 & 0.0116 & 0.87 & 0.0580 & 0.86 & 0.0116 & 0.84 & 0.0248 & 0.62 & 0.0802 & 0.80 & 0.0395 \\

        \bottomrule
    \end{tabular}
    }
    \caption{Average and variance results (of $5$ runs) for different methods. The best average values are shown in bold.}
    \label{tbl_pca_accy}
\end{table*}

\subsection{Discussion}
The proposed PCD2Vec achieved a 2\% improvement in average accuracy compared to the second-best baseline method. Since the baseline accuracy is above 90\% in most cases, a small increase in accuracy could have significant implications in the fight against the spread of the coronavirus, leading to early detection and containment, reducing false positives and unnecessary quarantines, and improving resource allocation. In conclusion, even a small 2\% improvement in performance can have a big impact on coronavirus host classification.

The class imbalance issue in the dataset was evaluated by measuring precision, recall, weighted F1-score, macro F1-score, and ROC-AUC of different models (using the one-vs.-rest approach). High scores of these metrics indicate good average performance across classes, and the results show that the class imbalance was handled effectively by all methods.

\subsection{Statistical Significance}
We also used the student t-test to evaluate if the classification results were statistically significant. We noted that the P-values were $<0.05$ in all cases, confirming the significance of the results. We have not reported the P-values in this paper due to limited space.

\section{Conclusion and Future Work} \label{sec_conclusion}
In this paper, we presented a novel method for predicting the host specificity of coronaviruses by analyzing spike protein sequences. Our method involves the use of Poisson correction distance, radial basis
function kernel, and kernel PCA to generate low-dimensional embeddings of the spike protein
sequences. Our method was tested on a real-world dataset of spike
protein sequences from different coronavirus species, and our results
showed that it outperforms other existing methods in terms of
predictive accuracy.
Our method has a solid theoretical foundation as we have provided
proofs for the three important properties of a distance matrix
(triangle inequality, symmetry, and non-negativity) which can be used
to justify its use in a machine-learning setting.  Our results
demonstrate the potential of our method to be used as a tool for
predicting the host specificity of coronaviruses, a critical step in
understanding the distribution, evolution, and spread of these
diseases.  In conclusion, our proposed method provides an effective, efficient, and reliable way to predict the host specificity of
coronaviruses by analyzing spike protein sequences. Future work will
focus on refining and improving our method, as well as testing it on
larger and more diverse datasets.

\bibliographystyle{plain}
\bibliography{references}

\end{document}